\documentclass[superscriptaddress,amsmath, amssymb, amsfonts, twocolumn, footinbib]{revtex4-2}
\usepackage[german,american,english]{babel}
\usepackage{graphicx}% Include figure files
\usepackage{graphics}% Include figure files
\usepackage{dcolumn}% Align table columns on decimal point
\usepackage{multirow}
\usepackage{bm}% bold math
\usepackage{latexsym}
\usepackage{amssymb}
\usepackage{amsmath}
\usepackage{amsfonts}
\usepackage{layout}
\usepackage{verbatim}
\usepackage{epsfig}
\usepackage{graphicx}
\usepackage{amsbsy}
\usepackage{lipsum}
\usepackage{cleveref}
\usepackage[toc]{appendix}
\usepackage{xcolor}

\newcommand{\bea}{\begin{eqnarray*}}
	\newcommand{\eea}{\end{eqnarray*}}
\newcommand{\bne}{\begin{equation*}}
\newcommand{\ede}{\end{equation*}}
\newcommand{\ba}{\arraycolsep 0.3ex \begin{array}{rl}}
\newcommand{\ea}{\end{array}}
\newcommand{\bnen}{\begin{equation}}
\newcommand{\eden}{\end{equation}}
\newcommand{\bean}{\begin{eqnarray}}
\newcommand{\eean}{\end{eqnarray}}
\newcommand{\bsen}{\begin{subequations}}
	\newcommand{\esen}{\end{subequations}}

\newcommand{\bna}{\begin{array}}
	\newcommand{\eda}{\end{array}}
\newcommand{\bnm}{\begin{enumerate}}
	\newcommand{\edm}{\end{enumerate}}

\newcommand {\pd} [2] {\frac{\partial #1}{\partial #2}}

\usepackage[utf8]{inputenc}

\begin{document}

\title{Spin-Hall effect in topological materials: Evaluating the proper spin current in systems with arbitrary degeneracies}
\author{Hongyang Ma}
\altaffiliation{These authors contributed equally to this work.}
\affiliation{School of Physics, The University of New South Wales, Sydney 2052, Australia}
\affiliation{Australian Research Council Centre of Excellence in Low-Energy Electronics Technologies, UNSW Node, The University of New South Wales, Sydney 2052, Australia}
\author{James H. Cullen}
\altaffiliation{These authors contributed equally to this work.}
\affiliation{School of Physics, The University of New South Wales, Sydney 2052, Australia}
\author{Serajum Monir}
\affiliation{School of Physics, The University of New South Wales, Sydney 2052, Australia}
\author{Rajib Rahman}
\affiliation{School of Physics, The University of New South Wales, Sydney 2052, Australia}
\author{Dimitrie Culcer}
\affiliation{School of Physics, The University of New South Wales, Sydney 2052, Australia}
\affiliation{Australian Research Council Centre of Excellence in Low-Energy Electronics Technologies, UNSW Node, The University of New South Wales, Sydney 2052, Australia}

\begin{abstract} 
The spin-Hall effect underpins some of the most active topics in modern physics, including spin torques and the inverse spin-Hall effect, yet it lacks a proper theoretical description. This makes it difficult to differentiate the SHE from other mechanisms, as well as differentiate band structure and disorder contributions. Here, by exploiting recent analytical breakthroughs in the understanding of the intrinsic spin-Hall effect, we devise a density functional theory method for evaluating the conserved (proper) spin current in a generic system. Spin non-conservation makes the conventional spin current physically meaningless, while the conserved spin current has been challenging to evaluate since it involves the position operator between Bloch bands. The novel method we introduce here can handle band structures with arbitrary degeneracies and incorporates all matrix elements of the position operator, including the notoriously challenging diagonal elements, which are associated with Fermi surface, group velocity, and dipolar effects but often diverge if not treated correctly. We apply this method to the most important classes of spin-Hall materials: topological insulators, 2D quantum spin-Hall insulators, non-collinear antiferromagnets, and strongly spin-orbit coupled metals. We demonstrate that the torque dipole systematically suppresses contributions to the conventional spin current such that, the proper spin current is generally smaller in magnitude and often has a different sign. Remarkably, its energy-dependence is relatively flat and featureless, and its magnitude is comparable in all classes of materials studied. These findings will guide the experiment in characterizing charge-to-spin interconversion in spintronic and orbitronic devices. We also discuss briefly a potential generalisation of the method to calculate extrinsic spin currents generated by disorder scattering.
\end{abstract}

\date{\today}

\maketitle

\section{Introduction}
Since its discovery two decades ago \cite{Exp-SHE-Science-2004, Exp-SHE-PRL-2005} the spin-Hall effect (SHE) has become one of the most actively studied topics in modern physics. Its uses range from the inverse SHE used as a detection and characterization tool \cite{Kimura-ISHE-PRL-2007,ISHE-Saitoh-Natcomm-2012,kimata2019magnetic,safeer2019room} to the generation of spin torques that flip magnetic bits \cite{Roadmap-SOT-Review, Manchon-RMP-2019}. The spin-Hall torque induces magnetization dynamics in spintronic memory devices \cite{ST-SHE-PRL,Ohno-Current-induced-Nat-Mat-2012,SOT-Review-APR} and is strong in many topological materials \cite{Manchon-Nat-2014, Control-SOT-Nat-Phys-2017, Strong-SOT-Nat-Com-2022, Hazra_NC2023,manchon2015new}, including, recently, van der Waals heterostructures coupled to WTe$_2$ \cite{SOT-WTe2}, Mn$_3$Sn \cite{Hazra_NC2023}, and heavy metals \cite{PhysRevApplied.14.064056}. Despite its manifold uses, the underlying mechanisms of the SHE remain somewhat mysterious and poorly understood. It remains difficult to distinguish experimentally between intrinsic (band structure) and extrinsic (disorder) contributions, as well as between the SHE and other mechanisms such as the orbital Hall and Edelstein effects \cite{SOT-Review-APR,Wang2018}. Hence, realistic calculations of the SHE for real materials are urgently needed.

The absence of an experimental blueprint for measuring the SHE accurately is related to the inherent difficulty in calculating the spin-Hall current. The conventional spin current is physically meaningless, since the spin precesses as it is transported \cite{Defination-SC-PRB-2005-XC,Sugimoto-CSC-PRB,Conserved-SC-PRB-2006,Universal-SC-Sci-Rep,Defintion-SC-PRL-2006-Qian,CSC-2D-hole-Qian-PRB-2008}. The proper spin current, based on the equation of continuity and the Onsager relations, is conserved, but contains a torque dipole term that involves the position operator \cite{Dimi-PRL-2004, SC-Shuichi-PRB-2004, Conserved-SC-Mott-PRB-Cong-2018, Defintion-SC-PRL-2006-Qian,CSC-2D-hole-Qian-PRB-2008,Cong-CC-PRB,Sugimoto-CSC-PRB,Hong_PSC,Cullen_extrinsic}. The intra-band elements of the position operator, which are associated with Fermi surface, group velocity, and dipolar effects, are challenging to evaluate in extended systems, since Bloch electrons are delocalized \cite{Sugimoto-CSC-PRB,Cong-CC-PRB,Hong_PSC,atencia2023non}. Additional major challenges include the presence of degeneracies in realistic band structures \cite{Hong_PSC,Cullen_extrinsic}, and the difficulty of bridging the gap between analytical approaches and density functional theory (DFT), while minimizing the computational cost.

In this work we exploit recent theoretical breakthroughs in evaluating the proper spin-Hall current in extended systems to overcome these challenges. We develop a new tight-binding approximation (TBA) methodology to determine the intrinsic proper spin current (IPSC), using the physical definition of Ref.~\cite{PhysRevLett.96.076604}, for different classes of topological materials as well as for strongly spin-orbit coupled metals. This paper breaks new ground: we develop a method for treating \textit{arbitrary matrix elements} of the position operator, including its diagonal elements, for Bloch electrons with \textit{arbitrary degeneracies}, and merge the analytical formalism with the TBA approach to yield a blueprint for evaluating the proper SHE for arbitrary band structures. We determine the size and structure of the spin current and its implications for the spin-Hall torque in topological materials. Of the materials we studied, Pt had the largest intrinsic spin current when using both the conventional and conserved definitions. Remarkably, in all classes of topological materials studied, as well as in metals, the proper SHE generally shares the same features. The torque dipole reduces the spin current and causes its energy-dependence to be relatively flat and featureless. This is in sharp contrast to the conventional spin current, which exhibits sharp peaks and dips as a function of energy. Using the proper definition significantly alters the calculated spin current, this cements the need to use the proper definition when making theoretical predictions. Furthermore, we find that the spin conductivities calculated using the proper definition for Pt and Mn$_3$Sn fit previous experimental results better.

Our work enables us to connect equilibrium density functional theory with non-equilibrium quantum mechanics based on the density matrix, which provides the most complete description of a quantum mechanical system. The results are presented in a form that is directly comparable to the experiment, where the SHE can be inferred from the spin-Hall angle. Whereas the focus here is on the intrinsic case, the method can be extended to disordered systems, incorporating skew scattering and side jump along the lines of Ref.~\onlinecite{Cullen_extrinsic}. The extension to disordered and inhomogeneous systems will enable the study of realistic devices and architectures in the most accurate and least computationally expensive way possible. 

This paper is organised as follows: first in section \uppercase\expandafter{\romannumeral2\relax} we discuss the proper spin current and present a general formula for its calculation. Next in section \uppercase\expandafter{\romannumeral3\relax} we present our computational methodology and results for the proper spin current and compare with results using the conventional spin current definition. We present results for 4 different materials; a topological insulator (Bi$_2$Se$_3$), 2D quantum spin-Hall insulator (WTe$_2$), antiferromagnet (Mn$_3$Sn), and a heavy metal (Pt). In section \uppercase\expandafter{\romannumeral4\relax} we discuss our calculated spin conductivities in the context of previous calculations and experimental results, we also discuss the utility of our new formula for spin-Hall effect calculations. Then in section \uppercase\expandafter{\romannumeral5\relax} we outline our methodology for defining the position operator in extended systems and offer a concise derivation of the formula for the appropriate spin current. And finally, we explore the feasibility of extending our calculations to systems characterized by disorder.

\section{Proper spin current}
Generating a spin current typically requires spin-orbit coupling, which causes spin precession and hence non-conservation. The conventional definition of the spin current is the product of the spin and velocity operators $J=\{\hat{v},\hat{s}\}$ \cite{Conventional-SC-Sci-Rep} or a redefined velocity operator \cite{Ghosh-PRB-2021}. However, the non-conservation of spin as it is transported makes the conventional spin current physically meaningless: its relationship to spin accumulation is not obvious \cite{SH-accumulation-Sinova-PRL-2005, Kleinert-2006}, it does not satisfy an equation of continuity or an Onsager relation, and is nonzero even in thermodynamic equilibrium \cite{Rashba-Spin-current-PRB, Defintion-SC-PRL-2006-Qian}. 

The proper spin current was introduced in Phys. Rev. Lett. {\bf{93}}, 046602 (2004) as a conserved spin current, which satisfies the equation of continuity. This led to the distinction between the conventional spin current, which is convenient but physically meaningless, and the proper spin current, which is conserved, despite being difficult to measure. The proper spin current operator includes the conventional spin current but with an extra contribution, the torque dipole $I=\{\hat{r},d\hat{s}/dt\}$ which arises from spin precession \cite{Dimi-PRL-2004, SC-Shuichi-PRB-2004, PhysRevLett.109.246604, Conserved-SC-Mott-PRB-Cong-2018, Defintion-SC-PRL-2006-Qian,CSC-2D-hole-Qian-PRB-2008,SHE-insulator-PRB-2020,Cong-CC-PRB}. Its calculation is subtle, involving matrix elements of the position operator between Bloch states. Operators containing the position operator are often difficult to deal with, as in a crystal in which the electron states are Bloch states the density matrix is diagonal in the crystal momentum. However, the position operator is not diagonal in the crystal momentum and couples wave vectors that are infinitesimally spaced.

The complications involved with properly defining the spin current can be avoided by calculating the spin response directly without resorting to the spin current \cite{Tatara-PRB-2018,Tatara-PRB-Letter}, which is appropriate when the quantity of interest is the spin accumulation. However, for phenomena that involve spin currents that don't have a spin accumulation such an approach is not appropriate and knowledge of the spin current is needed to gain physical insight. This is the case for spin-orbit torques driven by the spin-Hall effect or for the inverse spin-Hall effect. In spin-Hall torques, spin currents generated flow into a magnetic material and generate a torque on the magnetisation, in such a system there will not be any spin accumulation. In this context the calculation of spin currents is indispensable in interpreting spin torque experiments: it can reveal whether they are zero or finite, whether they change sign under certain circumstances, as well as their variation in different materials and with system parameters. The inverse spin-Hall effect is the Onsager inverse of the spin-Hall effect, it refers to the conversion of a spin current into a charge current. In this context, the spin-Hall conductivity, which is directly related to the spin current, is the quantity of interest. Hence, calculations of the spin current are crucial for the study of this effect.

Until now, there have been a number of works on the topic of calculating the proper spin current, each work has a distinct approach to the problem. In this work, which is an extension of Refs.~\onlinecite{Hong_PSC,Cullen_extrinsic}, we take an approach based on a quantum kinetic theory. Ref.~\onlinecite{Cong-CC-PRB} used a semiclassical approach based on wave packet dynamics. Lastly, Ref.~\onlinecite{Sugimoto-CSC-PRB} employed a Keldysh approach. There are important differences between each approach, and comparisons between results are not straightforward. 

Our approach is fully quantum mechanical and is equivalent to the Kubo linear response formalism, we evaluate the expectation value of the proper spin current operator operator. Whereas, the formalism in Ref.~\onlinecite{Cong-CC-PRB} constructs the proper spin current by adding a number of distinct contributions. These are formulated using classical physics and Maxwell’s equations while referencing the center of mass of a wave packet. Although the two approaches are quite different the final expressions for the proper spin current in a non-degenerate system are very similar, only differing by the position of spin and Berry connection matrix elements. However, it should be noted this work extends the results of Ref.~\onlinecite{Hong_PSC} by extending the formula to systems with degenerate states. In this work, we separate the torque dipole into two contributions $I_1$ and $I_2$, which represent the contributions from the degenerate/band-diagonal and non-degenerate matrix elements of the spin operator respectively.

The Keldysh approach of Ref.~\onlinecite{Sugimoto-CSC-PRB} calculates the torque dipole by employing a fictitious electromagnetic field. The calculation is semiclassical in the sense that it mixes both position and momentum coordinates. Their results find that the spin current vanishes in spin-1/2 systems regardless of the model. This differs from our findings and those of Ref.~\onlinecite{Cong-CC-PRB}. A detailed comparison is hampered by the lack of generic formulas in Ref.~\onlinecite{Sugimoto-CSC-PRB}. Furthermore, it is unclear what the analogues to the fictitious electromagnetic fields and explicit gradient expansion are in our approach.

In this work, we extend the result of Ref.~\onlinecite{Hong_PSC} for the intrinsic proper spin current to systems with arbitrary degeneracies, whereas the formula presented in this previous work is only valid in fully non-degenerate systems. Our general analytical expression for the IPSC spin-Hall conductivity in systems with arbitrary degeneracies is
\begin{equation}\label{CS-main}
    \sigma^l_{ij} = -\frac{2 e E_j}{\hbar}\sum_{{\bm k}} \sum_{mnn^\prime} f_{m{\bm k}} \text{Im}\left[\Tilde{\mathcal R}^{i}_{mn\bm k}\check{s}^{l}_{nn^\prime\bm k}\Tilde{\mathcal R}^{j}_{n^\prime m\bm k}\right],
\end{equation}
where $m$, $n$ and $n^\prime$ are band indices, ${\bm E}$ is the external electric field, $f_{m{\bm k}} \equiv f(\varepsilon_{{m\bm k}})$ the equilibrium Fermi-Dirac distribution, $\varepsilon_{m{\bm k}}$ the band dispersion, ${\bm k}$ the wave vector, ${\bm s}$ the spin operator, and ${\bf \mathcal R}_{mn\bm k}=\langle u_{m{\bm k}}|\mathrm{i}\pd{u_{n{\bm k}}}{\bm k}\rangle$ the Berry connection. The check over the spin term indicates band diagonal matrix elements and elements between degenerate states, and the tilde over the Berry connection indicates matrix elements between non-degenerate states. This is the \textit{only} intrinsic contribution to the proper spin current, which is shown to flow perpendicular to the applied electric field. Since only the band off-diagonal matrix elements of the Berry connection are included in the calculation the proper spin current is gauge invariant. It can also be finite in the insulating gap, with important implications for TI/FM devices, as we show below.

\section{Materials}
We consider 4 different types of materials in our calculations: a topological insulator (Bi$_2$Se$_3$), 2D quantum spin-Hall insulator (WTe$_2$), antiferromagnet (Mn$_3$Sn) and a heavy metal (Pt). We chose these materials due to their prevalence in spintronic research.\cite{han2018quantum} We employ density functional theory (DFT) to calculate the electronic and spintronic properties of Bi$_2$Se$_3$, WTe$_2$, Mn$_3$Sn, and Pt.

Our computational approach involves the utilization of the projector augmented-wave (PAW) method implemented within the Vienna \textit{ab-initio} Simulation Package (VASP) \cite{kresse_efficient_1996}. The Perdew–Burke–Ernzerhof (PBE) exchange-correlation functional \cite{perdew_generalized_1996} was employed, accompanied by standard scalar-relativistic PAW pseudopotentials \cite{blochl_projector_1994,kresse_ultrasoft_1999}. The electronic configurations for the valence electrons of Bi, Se, W, Te, Mn, Sn, and Pt were specified as $5d^{10}6s^26p^3$, $6s^26p^4$, $5s^25p^65d^{4}6s^2$, $5s^25p^4$, $3p^63d^{5}4s^2$, $4d^{10}5s^25p^2$, and $5d^96s^1$, respectively. The experimental crystal parameters for Bi$_2$Se$_3$ \cite{nakajima_crystal_1963}, WTe$_2$ \cite{brown_crystal_1966}, Mn$_3$Sn \cite{singh_study_1968}, and Pt \cite{owen_xli_1933} were adopted, with dimensions set as $a = b = 4.14$ and $c = 28.64$ Å; $a = 6.28$, $b = 3.50$, $c = 19.15$ Å; $a = b = 5.67$ and $c = 4.53$ Å; and $a = b = c = 3.92$ Å, respectively. For WTe$_2$, a Hubbard-like correction was applied within the Dudarev scheme \cite{dudarev1998electron}, with a U value of 5.1 eV chosen to accurately reproduce the correct band structure. Furthermore, all computations incorporated spin-orbit coupling (SOC), with a plane-wave basis set cutoff energy established at 350 eV. Convergence criteria for energy optimization were set to $10^{-6}$ eV. The Monkhorst–Pack ${\bm k}$-point grid was applied, with the ${\bm k}$-mesh tailored for different materials.

The determination of spin-Hall conductivity (SHC) entailed the use of a tight-binding Hamiltonian, derived from localized Wannier functions (WFs) \cite{marzari_maximally_2012,qiao_calculation_2018} projected from the DFT Bloch wavefunctions. Specifically, SHC calculations were performed through Qiao’s method \cite{qiao_calculation_2018}. Subsequent postprocessing was conducted using Wannier90 \cite{pizzi_wannier90_2020} and WannierTools \cite{wu2018wanniertools} codes.  Notably, atomic-orbital-like WFs encompassing Bi-p/Se-p, W-d/Te-p, Mn-p/Sn-p, and Pt-spd orbitals were selectively chosen for this analysis. The calculated band structures are in good agreement with \textit{ab-initio} results, plots of the band structures can be found in the supplemental material. 

\subsection{Bi$_2$Se$_3$}
DFT first-principles calculations were performed to analyze the band structure and spin-Hall conductivity (SHC) of Bi$_2$Se$_3$, a well-known topological insulator \cite{Hasan_review}. The crystal structure of Bi$_2$Se$_3$ is defined by the space group \textit{R$\overline{3}$m} and associated with the Laue group $\overline{3}$\textit{m}11$'$. We defined an energy range for the Wannier functions (WFs), spanning an inner frozen window from -5.8 to 1.8 eV and an outer disentanglement window from -5.8 to 14.2 eV relative to the Fermi level. This approach yielded 30 spinor WFs, exhibiting \textit{p}-like characteristics, which were used to construct a tight-binding Hamiltonian that replicates the \textit{ab-initio} band structure with high fidelity, as illustrated in Fig.~S1 and corroborated by the work of Ref.~\onlinecite{zhang2009topological}.

\begin{figure}[htbp]
\centering
\includegraphics[width = \linewidth]{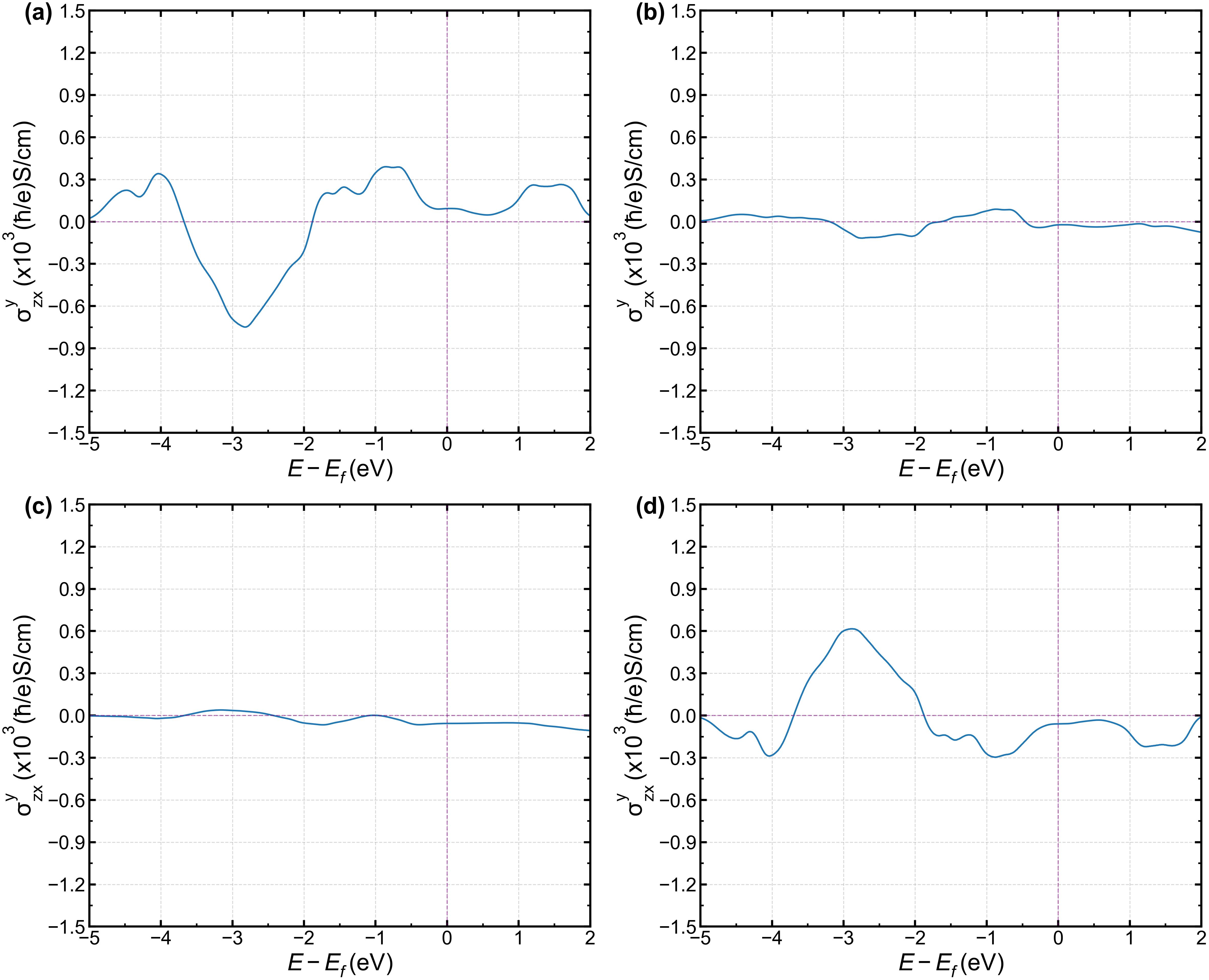}
\caption{spin-Hall Conductivity (SHC) $\sigma_{z x}^y$ vs energy for Bi$_2$Se$_3$. Panel (a) shows the SHC calculated using the conventional spin current formula, while panel (b) shows the SHC calculated using the proper spin current formula. Panels (c) and (d) contain the two torque dipole contributions $I_1$ and $I_2$, respectively.}
\label{fig:Bi2Se3-zxy}
\end{figure}

In Fig.~\ref{fig:Bi2Se3-zxy} we plot the SHC $\sigma_{z x}^y$ vs energy. The conventional SHC value at the Fermi level is calculated to be $93(\hbar/e)$S/cm, decreasing to $-749(\hbar/e)$S/cm at 2.82 eV below the Fermi level. The similarity between Fig.~\ref{fig:Bi2Se3-zxy}(a) and previously presented results \cite{PhysRevMaterials.4.114202} validates the accuracy of our calculation. The introduction of the torque dipole corrections, as shown in Fig.~\ref{fig:Bi2Se3-zxy}(b), significantly alters the SHC spectrum, mainly flattening it due to $I_2$ cancelling with the conventional spin current. We find the proper SHC $\sigma_{z x}^y$ to be $-22(\hbar/e)$S/cm at the Fermi level. Additionally, the introduction of the torque dipole has shown a similar effect on another spin conductivity tensor component, $\sigma_{x y}^z$, where the flattening of the spin conductivities energy dependence is also seen; however, the proper SHC value at the Fermi level increases to $165(\hbar/e)$S/cm, contrasting with the $80(\hbar/e)$S/cm determined through the conventional spin current.

\subsection{WTe$_2$}
The second material that we investigated was 1T$'$-WTe$_2$, a monolayer 2D quantum spin-Hall insulator distinguished by its stability and electronic properties among 1T$'$ phase monolayer transition metal dichalcogenides \cite{tang2017quantum}. The choice of WTe$_2$ was motivated by its promising potential in spintronic and orbitronic devices.  WTe$_2$’s crystal structure is defined by the space group \textit{Pm} and classified within the Laue group 2$'$/\textit{m}$'$. Through a meticulously chosen energy window we extracted 44 spinor WFs, we used an inner frozen window spanning from -7.2 to 2.5 eV and an outer disentanglement window ranging from -7.2 to 9.3 eV relative to the Fermi level. Our spinor WFs have W-\textit{d}/Te-\textit{p}-like Gaussian forms and were used to construct a tight-binding Hamiltonian that precisely mirrors the \textit{ab-initio} band structure, as shown in Fig.~S1.

\begin{figure}[t]
\centering
\includegraphics[width=\linewidth]{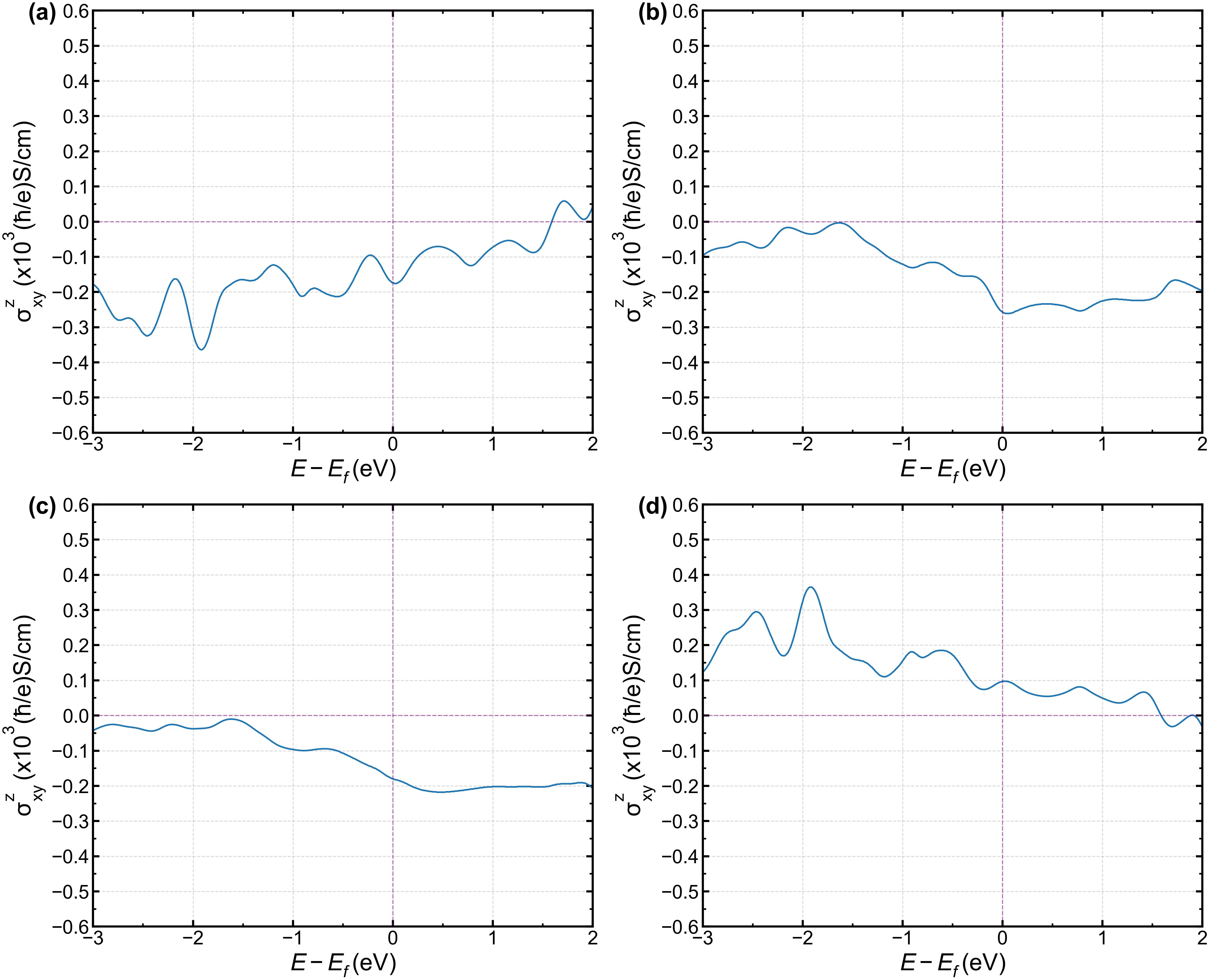}
\caption{spin-Hall Conductivity (SHC) $\sigma_{x y}^z$ vs energy for WTe$_2$. Panel (a) shows the SHC calculated using the conventional spin current formula, while panel (b) shows the SHC calculated using the proper spin current formula. Panels (c) and (d) contain the two torque dipole contributions $I_1$ and $I_2$, respectively.}
\label{fig:WTe2-xyz}
\end{figure}

Our calculated conventional SHC for WTe$_2$, $\sigma_{x y}^z$, displayed in Fig.~\ref{fig:WTe2-xyz}, shows strong agreement with previously reported values \cite{PhysRevB.99.060408}, with the conventional SHC at the Fermi level calculated to be $-174(\hbar/e)$S/cm, increasing in magnitude to $-364(\hbar/e)$S/cm at 1.92 eV below the Fermi level. The inclusion of the torque dipole corrections, as depicted in Fig.~\ref{fig:WTe2-xyz}(b), substantially modifies the SHC spectrum. This is again primarily due to $I_2$ canceling with contribution from the conventional spin current. Using the proper spin current definition we find $\sigma_{x y}^z$ to be $-257(\hbar/e)$S/cm at the Fermi level. Additionally, we calculated $\sigma_{z x}^y$ which exhibited a similar cancellation due to the introduction of $I_2$. However, for this tensor component, the proper SHC value at the Fermi level exhibited negligible variation from the conventional definition adjusting to $10(\hbar/e)$S/cm from $9(\hbar/e)$S/cm.

\subsection{Mn$_3$Sn}
We then calculated the spin conductivity of Mn$_3$Sn, a non-collinear antiferromagnetic material at room temperature \cite{tomiyoshi1982,nakatsuji2015}. This material's crystal structure is classified by the space group \textit{P6$_3$/mmc} and the Laue group 6/\textit{mmm}. The derivation of 72 spinor WFs, with Mn-\textit{d}/Sn-\textit{p}-like Gaussian forms, was done using inner frozen and outer disentanglement windows spanning from -7.1 to 0.7 eV and -7.1 to 15.1 eV relative to the Fermi level, respectively. We again used the WFs to construct a tight binding model that accurately replicated the \textit{ab-initio} band structure (Fig.~S1) \cite{li2023field}, demonstrating the accuracy of our tight-binding model.

\begin{figure}[t]
\centering
\includegraphics[width=\linewidth]{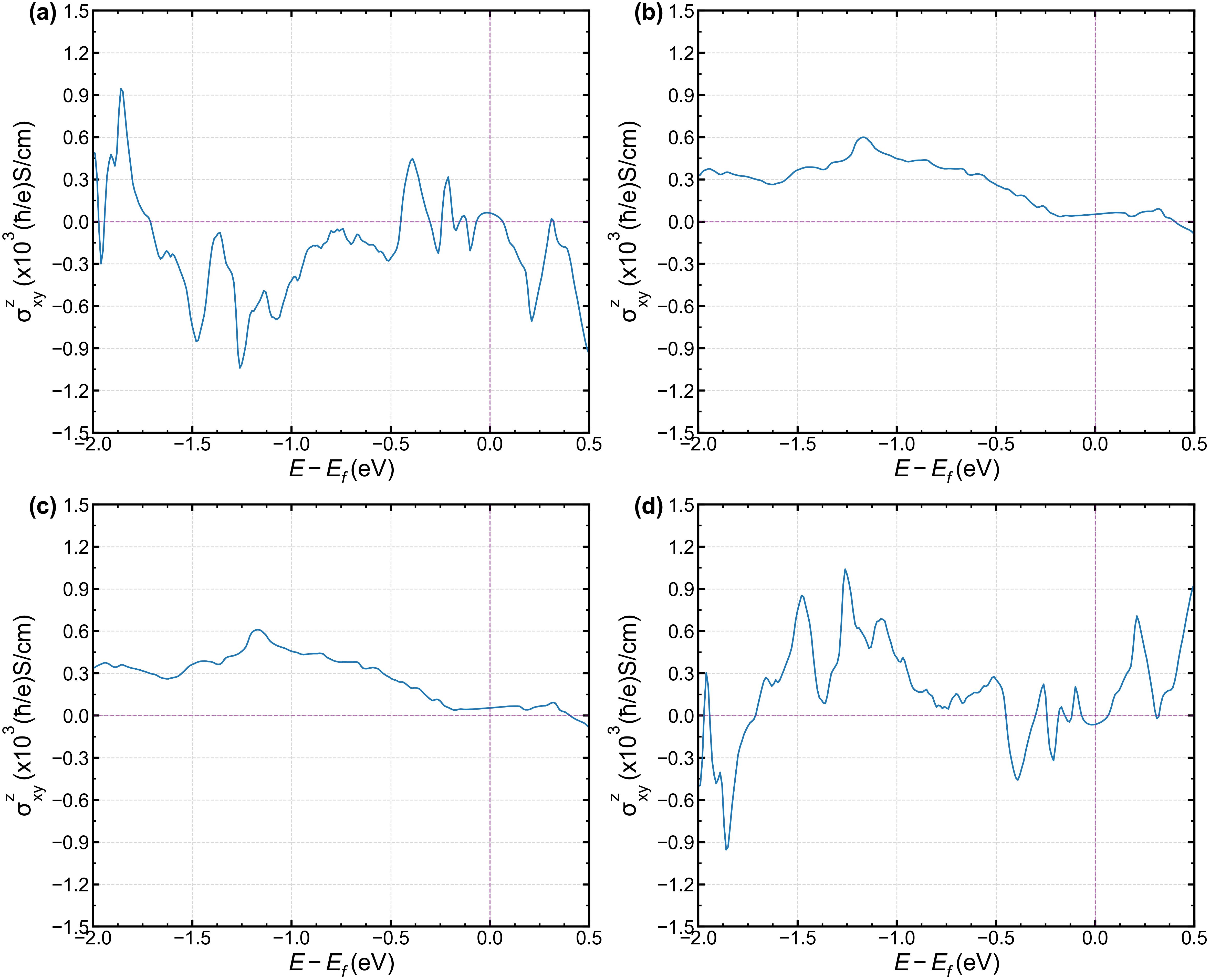}
\caption{spin-Hall Conductivity (SHC) $\sigma_{x y}^z$ vs energy for Mn$_3$Sn. Panel (a) shows the SHC calculated using the conventional spin current formula, while panel (b) shows the SHC calculated using the proper spin current formula. Panels (c) and (d) contain the two torque dipole contributions $I_1$ and $I_2$, respectively.}
\label{fig:Mn3Sn-xyz}
\end{figure}

As depicted in Fig.~\ref{fig:Mn3Sn-xyz}, conventional SHC at the Fermi level exhibits sharp variation across different energy levels. At the Fermi level, the conventional SHC is calculated to be $61(\hbar/e)$S/cm changing to $-1044(\hbar/e)$S/cm at 1.26 eV below the Fermi level. Our results for the conventional SHC are consistent with previously published data \cite{zhang2017strong}. The inclusion of the torque dipole significantly alters the SHC spectrum, as illustrated in Fig.~\ref{fig:Mn3Sn-xyz}(b), flattening out the sharp energy dependence. Among the materials studied, Mn$_3$Sn seems to have the most profound correction from $I_2$, it almost completely cancels with the conventional spin current. Hence, $I_1$ is the primary factor in determining the proper SHC spectrum, this is clear when comparing Fig.~\ref{fig:Mn3Sn-xyz}(b) and (c). We find the proper SHC $\sigma_{x y}^z$ at the Fermi level to be $53(\hbar/e)$S/cm. Furthermore, another component of the spin conductivity tensor, $\sigma_{z x}^y$, has similar cancellation between $I_2$ and the conventional spin current. We find the value of $\sigma_{z x}^y$ to be $47(\hbar/e)$S/cm using the proper definition as opposed to $-91(\hbar/e)$S/cm using the conventional definition.

\subsection{Pt}
The last material we considered was Pt, a heavy metal with a significant role in spintronics due to its strong spin-orbit coupling \cite{Liu2011,Roadmap-SOT-Review}. The crystal structure of Pt is classified by the space group \textit{Fm$\overline{3}$m} and the Laue group \textit{m}$\overline{3}$\textit{m}. We derived 18 spinor WFs, which exhibit Pt-\textit{spd}-like Gaussian forms, using an inner frozen window and an outer disentanglement window, spanning from -11 to 6.7 eV and  -11 to 33.7 eV relative to the Fermi energy, respectively. We again used the WFs to create a tight binding model that replicates the \textit{ab-initio} band structure accurately, as depicted in Fig.~S1.

\begin{figure}[t]
\centering
\includegraphics[width=\linewidth]{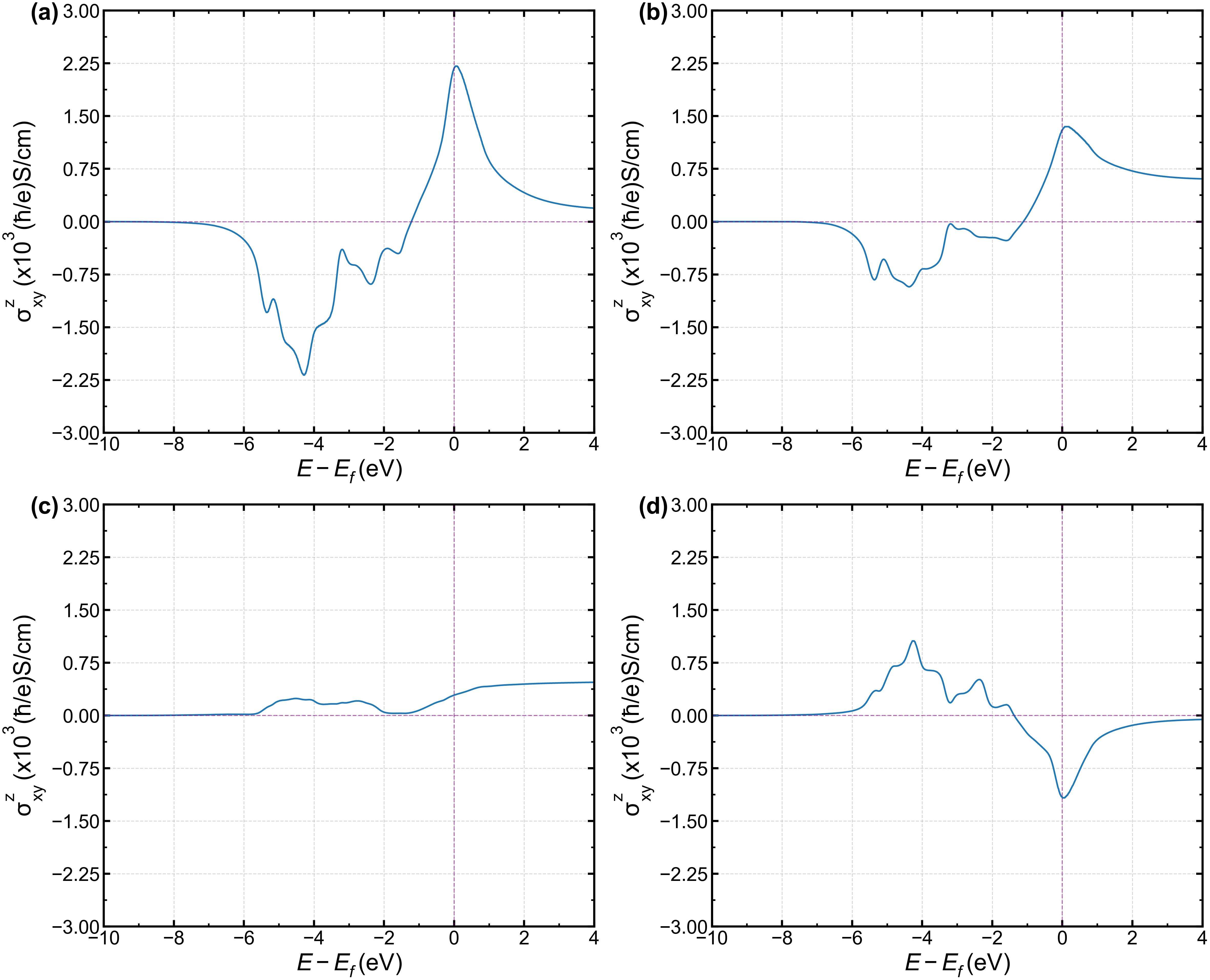}
\caption{spin-Hall Conductivity (SHC) $\sigma_{x y}^z$ vs energy for Pt. Panel (a) shows the SHC calculated using the conventional spin current formula, while panel (b) shows the SHC calculated using the proper spin current formula. Panels (c) and (d) contain the two torque dipole contributions $I_1$ and $I_2$, respectively.}
\label{fig:Pt-xyz}
\end{figure}

Fig.~\ref{fig:Pt-xyz} plots the SHC, $\sigma_{x y}^z$, at different energy levels. At the Fermi level, the conventional SHC reaches $2180(\hbar/e)$S/cm declining and changing the sign to a value of $-2177(\hbar/e)$S/cm at 4.28 eV below the Fermi level. Our calculation of the conventional SHC aligns with previous calculations ~\cite{qiao_calculation_2018}. The inclusion of the torque dipole results in a proper SHC $\sigma_{x y}^z$ of $1305(\hbar/e)$S/cm at the Fermi level. The energy dependence of the proper SHC, depicted in Fig.~\ref{fig:Pt-xyz}(b), shows that the torque dipole introduces a small but non-negligible correction to the SHC in Pt, differing from the more pronounced effects observed in materials like Mn$_3$Sn. This highlights the nuanced role these corrections play in the overall spectral behavior.

\section{Discussion}
In general, our spin conductivity results show that the magnitude of the proper spin current will generally be smaller than the conventional spin current. This is important as it highlights that previous theoretical works employing the conventional definition may overestimate the size of the spin-Hall effect. Furthermore, the spin conductivities found using the proper definition for Pt and Mn$_3$Sn fit experimental results better than the those found using the conventional definition. For the spin conductivity calculations in the other materials neither the conventional nor the proper definition fits the experimental results, we expect that this is due to effects other than the spin-Hall effect dominating in these measurements. Our figures also show that the energy dependence of the proper spin current is flatter than the conventional spin current, which seems to have sharp fluctuations that the proper spin current does not. The reduction in magnitude and flatter energy dependence is due to the torque dipole cancelling exactly with terms in the conventional spin current. This is prominently seen in Figs.~\labelcref{fig:Bi2Se3-zxy,fig:WTe2-xyz,fig:Mn3Sn-xyz,fig:Pt-xyz} when comparing the second part of the torque dipole with the conventional spin current. 

Our results for the spin conductivity in Bi$_2$Se$_3$ for $\sigma^y_{zx}$ predict the spin conductivity to have a different sign to previous results using a $\boldsymbol{k}\cdot\boldsymbol{p}$ Hamiltonian. These previous results found the spin conductivity to be $8(\hbar/e)$S/cm \cite{Hong_PSC}, whereas we find the spin conductivity to be $\sigma^y_{zx}=-22(\hbar/e)$S/cm. This difference could be due to previous calculations employing a Hamiltonian that had a magnetisation, which was not considered in this work. Nonetheless, our results remain two orders of magnitude lower than the larger spin torque measurements observed in Bi$_2$Se$_3$ devices \cite{Mellnik2014,Dc2018}, hence, effects such as the spin-transfer torque and Rashba-Edelstein effect are likely responsible for these larger torques \cite{cullen2023spin}. The spin conductivity $\sigma^y_{zx}$ has opposite sign when calculated using the proper definition. Hence, spins generated via the SHE have opposite sign to spins generated via the REE \cite{sakai2014, Fischer2016,farokhnezhad2022spin,farokhnezhad2023spin}, this is consistent with the experimental results that find the spin torque efficiency increase for thinner TI samples where the SHE is smaller \cite{Wang2017}.

A spin-orbit torque experiment recorded an out-of-plane field-like torque conductivity in WTe$_2$ of $45(\hbar/e)$S/cm \cite{macneill2017control}. This number can be compared with our results for $\sigma^y_{zx}$. Comparing the experimental result with our result of $10(\hbar/e)$S/cm, our calculated spin conductivity is the same order of magnitude, though the experimental result is substantially larger. However, it should be noted that these experimental results are for bilayer WTe$_2$, whereas our results are for a monolayer structure. Another study of spin torques in WTe$_2$ devices found the spin conductivity to be up to two orders of magnitude larger \cite{li2018spin}, however, it ascribes this large torque to surface state effects. Furthermore, the spin conductivity calculated using the conventional definition is of a very similar magnitude $\sigma^y_{zx}=9(\hbar/e)$S/cm, and hence does not fit the experimental results any better.

Our results in Mn$_3$Sn, shown in Fig.~\ref{fig:Mn3Sn-xyz}, find the magnitude of proper spin conductivity at the Fermi level $\sigma^z_{xy}=53(\hbar/e)$S/cm to be smaller than the conventional spin conductivity $\sigma^z_{xy}=61(\hbar/e)$S/cm. Our results for the energy dependence of the spin conductivity using the conventional definition agree with previous calculations \cite{zhang2017strong,guo2017large,zhang2018spin}. 
The spin conductivity in Mn$_3$Sn has been measured to be $\sigma_{\text{SH}}=46.99\pm20.63(\hbar/e)$S/cm in an ISHE experiment \cite{muduli2019evaluation}. Our proper spin current results agree very well with the magnitude of the measured spin conductivity, and our results have the same sign as the experiment. The conventional spin conductivity result also agrees with this experimental result, though it falls toward the upper end of the range.

Analysis of spin-Hall effect measurements in Pt finds its spin conductivity to be $(0.7-1.7)\times 10^3(\hbar/e)$S/cm \cite{liu2011review}. These values are in very good agreement with our results using the proper spin current definition, as is shown in Fig.~\ref{fig:Pt-xyz}. At the Fermi energy we find the spin conductivity to be $\sigma^z_{xy}=1.31\times10^3(\hbar/e)$S/cm. The conventional spin current result of $\sigma^z_{xy}=2.18\times10^3(\hbar/e)$S/cm also loosely agrees with these measurements, however, it falls just above the upper end of the range.

The spin current formula we present here extends the work in Ref.~\onlinecite{Hong_PSC} to systems with arbitrary degeneracies. Whereas, the previous formula was only valid in fully non-degenerate systems. Furthermore, in this work we have demonstrated that this method for calculating the proper spin current is not restricted to $\boldsymbol{k}\cdot\boldsymbol{p}$ models and can be straightforwardly used with DFT calculations. The use of more accurate models that are accurate beyond the band center is necessary for spin current calculations, since all wavevectors in the Brioullin zone must be summed over and filled bands can generate nonzero spin currents.

For the proper spin current to be strictly conserved, the expectation value of the torque, ${\rm Tr} (\hat{t}\hat{\rho})$, needs to cancel so that globally there is no net spin generation in the system \cite{Dimi-PRL-2004, Defintion-SC-PRL-2006-Qian}. In other words the torque density ${\rm Tr} (\hat{t}\hat{\rho})$ vanishes but the torque dipole density ${\rm Tr} (\{\hat{t}, \hat{{\bm r}}\} \hat{\rho})$ is finite. This is true for the models we consider in this paper. % Is it actually true for all these models?

It is well known that the spin accumulation depends crucially on boundary conditions \cite{QSHE-SOC-PRL-2006-Shuichi,Defintion-SC-PRL-2006-Qian, SHE-insulator-PRB-2020, PhysRevB.72.241303}, and Ref.~\cite{Tatara-PRB-Letter} demonstrated quantitatively that the spin accumulation can be determined without reference to the spin current. Hence, in systems where spin accumulation is the quantity of interest direct calculation of the spin accumulation without reference to the spin current can be favourable. However, accurately defining the boundary conditions, which are often unknown, is a limiting factor for such calculations.

Our calculation is indispensable in systems in which the spin current does not lead to a spin accumulation. Such systems, which include TI/FM interfaces, are in fact used to infer the presence of a spin current. Since the spin current does not couple to any measurable quantity, its detection is primarily through indirect processes, for example by measuring spin-torque driven magnetization precession \cite{SC-indirect-measure-PRL,SC-indirect-Nat-SOT}, spin-current induced second-harmonic optical effects \cite{SC-SHM-indirect,SC-indirect-SHM-Nat}, the inverse spin-Hall effect \cite{Kimura-ISHE-PRL-2007,ISHE-Saitoh-Natcomm-2012,kimata2019magnetic}, and X-ray pump-probe measurements \cite{SC-Pump_probe-PRL}.

\section{Derivation of the proper spin current}

We outline in this section the derivation leading to Eq. 1. We first discuss our methodology for dealing with arbitrary matrix elements of the position operator, which is vital in obtaining the correct expression for the torque dipole. Next we apply this methodology to determine the full expression for the proper spin current in systems with arbitrary degeneracies. Finally, we discuss briefly the procedure for extending this formalism to disordered systems.   

\subsection{The position operator between Bloch states}

For a system described by a single-particle density operator $\hat{\rho}$, the expectation value of an arbitrary operator $\hat{O}$ is given by $\langle \hat{O} \rangle = \text{Tr}\,\hat{O}\hat{\rho}$. Operators containing the position operator can be difficult to deal with, as in a crystal in which the electron states are Bloch states the density matrix $\hat{\rho}$ is diagonal in the crystal momentum. Whereas the matrix elements of the position operator in the crystal momentum representation couple wave vectors that are infinitesimally spaced. Here we outline the derivation of a general expression for the trace of an operator with the position operator. We consider the trace of the position operator with some operator $\hat{\Lambda}$,
\begin{equation}\label{eq:Tr}
\begin{aligned}
    \frac{1}{2}\text{Tr}\{\hat{r}_i,\hat{\Lambda}\}=\frac{1}{2}\sum& \langle \psi_{n \bm k}|\hat{r}_j|\psi_{m \bm k^\prime}\rangle\langle\psi_{m \bm k^\prime}|\hat{\Lambda}|\psi_{n \bm k}\rangle+\\
    &\langle \psi_{n \bm k}|\hat{\Lambda}|\psi_{m \bm k^\prime}\rangle\langle\psi_{m \bm k^\prime}|\hat{r}_j|\psi_{n \bm k}\rangle\,.
\end{aligned}
\end{equation}
Since the wavefunctions are Bloch states we can express them as $|\psi_{n \bm k}\rangle = e^{i\bm k \cdot \bm r}|u_{m \bm k}\rangle$. Therefore the position operator can be expressed as
\begin{equation}
    \langle \psi_{n \bm k^\prime}|\hat{r}_j|\psi_{m \bm k}\rangle=\langle u_{n \bm k^\prime}|e^{-i\bm k^\prime \cdot \bm r}\left(-i\pd{}{k_j} e^{i\bm k \cdot \bm r}\right)|u_{m \bm k}\rangle\,.
\end{equation}
Evaluating (\ref{eq:Tr}) and making the substitution $\bm k = \bm k_+$ and $\bm k^\prime = \bm k_-$ where $\bm k_{\pm} = \bm k \pm \bm Q/2$ we find that
\begin{equation}\label{eq:DipoleTr}
    \frac{1}{2}\text{Tr}\{\hat{r}_i,\hat{\Lambda}\}=\text{Tr}\mathcal{D}\{\hat{\Lambda}\}\,
\end{equation}
where $\mathcal{D}$ is a covariant derivative defined as \cite{atencia2023non}
\begin{equation}\label{eq:cd}
   \mathcal{D}\{\Lambda\}_{\boldsymbol{k}} \equiv \frac{1}{2}\left[\mathrm{i} \, \bigg(\pd{\Lambda_{{\bm k}_+ {\bm k}_-}}{\bm Q} - \pd{\Lambda_{{\bm k}_- {\bm k}_+}}{\bm Q}\bigg)_{{\bm Q} \rightarrow 0} + \{ \bm{\mathcal R}_{\bm k}, \Lambda_{\bm k} \}\right]\,.
\end{equation}
The wavevector off-diagonal nature of the position operator is taken care of by the substitution of the infinitesimal wavevector $\bm Q$. The differential terms in (\ref{eq:cd}) represent the phase of the wave function in more conventional evaluations, while the Berry connection represents the contribution due to the change of the basis states between infinitesimally-separated wave vectors.

In (\ref{eq:cd}) we consider density matrix elements that are infinitesimally off-diagonal in wavevector, whereas we consider the Berry connection to be purely diagonal in wavevector. This is because the Berry connection is defined as
\begin{equation}
    \mathcal{R}_{mn,\boldsymbol{k}\boldsymbol{k}}=\langle u_{m\boldsymbol{k}}| i \nabla u_{n \boldsymbol{k}}\rangle\,,
\end{equation}
and definitionally cannot have elements off-diagonal in the wave vector. It is straightforward to show that all formulas are gauge invariant. 

\subsection{Proper spin current in degenerate bands}

The system is described by a single-particle density operator $\hat{\rho}$, which obeys the quantum Liouville equation
\begin{equation}\label{QLE}
\pd{\hat{\rho}}{t}+\frac{\mathrm{i}}{\hbar}[\hat{H},\hat{\rho}]=0,
\end{equation}
where $\hat{H}$ is the total Hamiltonian of the system. We will consider an arbitrary Hamiltonian and focus on a clean system, deferring the treatment of disorder to a future publication. The conserved spin current operator $\hat{\mathcal{J}}^i_j = d/dt \, (\hat{r}_j\hat{s}_i)$ is the time derivative of the spin dipole operator. Taking trace with the single particle density matrix operator $\hat{\rho}$, we separate the conserved spin current into the conventional spin current and torque dipole contributions $ \mathcal{J}^i_j = \frac{1}{2} \, {\rm Tr} \, \hat{\rho} \, \{ \hat{s}_i, \hat{v}_j \} + \frac{1}{2} \, {\rm Tr} \, \hat{\rho} \, \{ \hat{t}_i, \hat{r}_j \}$, with the velocity operator $\hat{v}_j=d\hat{r}_j/dt$ and the torque $\hat{t}_i = d\hat{s}_i/dt$, both diagonal in wave vector in the crystal momentum representation. The conventional spin current $J^i_j =\frac{1}{2}{\rm Tr} \hat{s}_i \{ \hat{v}_j, \hat{\rho} \}$ is straightforward to evaluate. In contrast $I^i_j =\frac{1}{2} {\rm Tr} \hat{\rho} \{ \hat{t}_i, \hat{r}_j\}$ stemming from the torque requires some work to deal with the position operator $\hat{r}_j$. In order to deal with the position operator in the Bloch representation and evaluate the torque dipole we perform the manipulations outlined in the previous section, taking $\hat{\Lambda}=\hat{t}_i\hat{\rho}$ and evaluating (\ref{eq:DipoleTr}) we find
\begin{equation}\label{Ifin-main}
	 I^i_{j} = \mathrm{i} \, {\rm Tr} \, t_i \Xi_j.
\end{equation}
Where, $\boldsymbol{\Xi}_{\bm k}=\mathcal{D}\{\rho\}_{\boldsymbol{k}}$ \cite{atencia2023non}. It is easy to prove Eq.~(\ref{Ifin-main}) is gauge invariant. The torque $t_i=\frac{i}{\hbar}[H_0,s_i]$ is purely off-diagonal in band index. Hence, when evaluating the trace (\ref{Ifin-main}) only band off-diagonal elements of $\boldsymbol{\Xi}$ are required. To find $\boldsymbol{\Xi}$ the covariant derivative is applied to the quantum kinetic equation from Refs.~\onlinecite{Interband-coherence-PRB-2017,Rhonald-PR-Res-2022}, the resulting kinetic equation is
\begin{equation}\label{eq:qke}
    \displaystyle \pd{\boldsymbol{\Xi}_{\bm k}}{t} + \frac{i}{\hbar} \, [H_0, \boldsymbol{\Xi}_{\bm k}] = \displaystyle - \frac{i}{\hbar} \left[H_E,\boldsymbol{\Xi}_{\bm k}\right] - \frac{i}{2\hbar} \bigg\{\frac{DH_0}{D{\bm k}}, \rho_{\bm k} \bigg\}.
\end{equation}
Where the $\bm k$ diagonal density matrix $\rho_{\bm k}$ is found using the original kinetic equation. The new kinetic equation is solved for $\boldsymbol{\Xi}_{\bm k}$ in an identical manner, the details of this calculation can be found in the supplemental material.

The process of applying the covariant derivative to the kinetic equation and solving for $\boldsymbol{\Xi}_{\bm k}$ is equivalent to the approaches of Refs.~\onlinecite{Hong_PSC,Cullen_extrinsic}. In these works the quantum kinetic equation was expanded to linear order in the infinitesimal off-diagonal wavevector $\boldsymbol{Q}$ and then solved for $\rho_{\boldsymbol{k}_+ \boldsymbol{k}_-}$. The quantity $\boldsymbol{\Xi}$ is just the linear order correction to the density matrix in $\bm Q$, such that $\rho_{\bm k_+ \bm k_-}=\rho_{\bm k}+{\bm Q}\cdot\boldsymbol{\Xi}_{\bm k}$.  Furthermore, the expanded kinetic equation used in Refs.~\onlinecite{Hong_PSC,Cullen_extrinsic} is essentially identical to (\ref{eq:qke}). We would like to emphasize that the approach employed for the position operator in this work has the advantage of being more general and can be applied to evaluate any dipole operator in a homogeneous system \cite{atencia2023non}.

Solving (\ref{eq:qke}) and taking the trace (\ref{Ifin-main}) yield two contributions to the torque dipole. The first contribution is
\begin{equation}\label{eq:I1}
    I^i_{j,1}=\frac{i e \boldsymbol{E}}{2 \hbar}\cdot \sum_{m} f_m \left\{[\Tilde{\mathcal{R}}^j,\check{s}^i],\boldsymbol{\Tilde{\mathcal{R}}}\right\}_{mm},
\end{equation}
when expanding this expression half of the terms will exactly cancel with the conventional spin current, the other half will add to the spin current and yield the expression in (\ref{CS-main}). The second contribution to the torque dipole is
\begin{equation}\label{eq:I2}
\begin{aligned}
    \displaystyle I^i_{j,2} = \displaystyle -\frac{1}{2} \sum_{mn}& \,\frac{1}{\hbar} s^i_{mn} \bigg( \pd{\varepsilon_m}{{\bm k}} + \pd{\varepsilon_n}{{\bm k}} \bigg) \, S^{E}_{\bm k, {nm}}+\\
    &\frac{i}{\hbar} \, s^{i}_{mn} \{[\mathcal{R},H_0], S^{E}_{\bm k} \}_{nm},
\end{aligned}
\end{equation}
where $S^E_{\bm k}$ is the band off-diagonal part of the $\bm k$ diagonal non-equilibrium density matrix. All of the terms in (\ref{eq:I2}) will cancel exactly with the remaining contributions from the conventional spin current. Hence, the intrinsic contribution to the conventional spin current is contained in (\ref{CS-main}).

\subsection{Extension to Disordered Systems}

The next important step in gaining a complete understanding of the spin-Hall effect is to extend the theory to account for spin currents generated via extrinsic mechanisms. Our formalism can be straightforwardly extended to the case of disorder \cite{Interband-coherence-PRB-2017, Rhonald-PR-Res-2022, Culcer_SJ_PRB10}. A blueprint for the calculation of extrinsic spin currents has been presented in Ref.~\onlinecite{Cullen_extrinsic}, and a calculation was done with a $\boldsymbol{k}\cdot\boldsymbol{p}$ Hamiltonian. In this work it was shown that spin currents due to spin-orbit scattering effects such as skew and side jump scattering are of a similar order of magnitude to intrinsic spin currents. Furthermore, these contributions will appear at zeroth order in disorder, making them indistinguishable from intrinsic spin currents. This is consistent with previous results for the anomalous Hall effect \cite{sinitsyn2007semiclassical,nagaosa2010anomalous}, and for the conventional spin current \cite{Sugimoto-CSC-PRB, mishchenko2004spin}. Recently it was shown that extrinsic mechanisms in a similar effect, the orbital Hall effect, dominate the intrinsic orbital Hall effect \cite{liu2023dominance}, further highlighting the need to extend our theory to include disorder. The remaining challenge is in extending this formalism to \textit{ab-initio} and DFT calculations as was done in this work for intrinsic spin current. Including general disorder effects in such calculations is notoriously difficult. We suggest the use of the same simple disorder model as Refs.~\onlinecite{Cullen_extrinsic,Interband-coherence-PRB-2017,Rhonald-PR-Res-2022} and to calculate the disorder contribution iteratively as was done in Ref.~\onlinecite{Cullen_extrinsic} or by using a simple relaxation time approximation or Bloch lifetime.

\section{Conclusions} 
We have developed a method for the calculation of the proper spin current in systems with arbitrary degeneracies. We connected equillibrium density functional theory with our electromagnetic response theory based on the single particle density matrix to evaluate the proper spin current with realistic band structures. We calculated the proper spin current in a wide range of topological materials: Bi$_2$Se$_3$, WTe$_2$, Mn$_3$Sn, and Pt. Our method for the calculation of the proper spin current can be applied in future for further theoretical spin-Hall effect calculations in other materials. Furthermore, we demonstrate that the torque dipole directly cancels contributions to the conventional spin current, due to this the proper spin current is generally smaller in magnitude and can differ in sign from the conventional spin current. The large difference in results between the two definitions cements the need for the use of the proper definition in theoretical spin-Hall effect studies. Lastly, we briefly outlined how this theory may be extended to include extrinsic spin currents due to impurity scattering.

\section{Acknowledgments.} This work is supported by the Australian Research Council Centre of Excellence in Future Low-Energy Electronics Technologies, project number CE170100039. The research was undertaken with the assistance of resources and services from the National Computational Infrastructure (NCI) under the NCMAS 2022 \& 2023 allocation and the Research Technology Services at UNSW Sydney. JHC acknowledges support from an Australian Government Research Training Program (RTP) Scholarship. We are grateful to Binghai Yan, Tobias Holder, Daniel Kaplan, and Changming Yue.

\end{document}